# The Transient Optical Sky Survey Data Pipeline

E. Hadjiyska[1,2], G. Hughes[3], P. Lubin[1], S. Taylor[4], R. Hartong-Redden[1], J. Zierten[1]

[1] Department of Physics, University of California, Santa Barbara, CA 93106-9530, USA
[2] Yale University, Department of Physics, P.O. Box 208120, New Haven, CT 06520-8120, USA;
[3] California Polytechnic State University, San Luis Obispo, CA 93407, USA
[4] Raytheon Vision Systems, 75 Coromar Dr, Goleta, CA 93117, USA

ABSTRACT

The Transient Optical Sky Survey (TOSS) is an automated, ground-based telescope system dedicated to searching for optical transient events. Small telescope tubes are mounted on a tracking, semi-equatorial frame with a single polar axis. Each fixed-declination telescope records successive exposures which overlap in right ascension. Nightly observations produce time-series images of fixed fields within each declination band. We describe the TOSS data pipeline, including automated routines used for image calibration, object detection and identification, astrometry, and differential photometry. Time series of nightly observations are accumulated in a database for each declination band. Despite the modest cost of the mechanical system, results from the 2009-2010 observing campaign confirm the system's capability for producing light curves of satisfactory accuracy. Transients can be extracted from the individual time-series by identifying deviations from baseline variability.

*Online-only material:* color figures

1. INTRODUCTION

TOSS is a system of optical telescopes on a common mount (Figure 1) that seeks to produce automated nightly photometric assessments of objects within the system's observable fixed-declination band (Hadjiyska et al. 2008). The current TOSS configuration has a one day cadence, reaches depth of $m_v \approx 17$ in 90 s exposures and saturates at $m_v \approx 11$. TOSS is similar to existing automated transient detection schemes, but with several unique characteristics. For example, the All-Sky Automated Survey (ASAS) is also a low-cost transient survey configured for objects with $m_v < 13$ (Paczyński, 2000) and optimized for shorter-period events (Pojmański, 1997). TOSS revisits each field once nightly, seeking longer-period (>2 days) transient events.

TOSS is polar-aligned, with sidereal tracking for recording long-exposure images. Commercially available color cameras are mounted to the base of each telescope. Images are recorded by starting each exposure with the leading telescope pointing at zenith. The mount tracks the sky for the duration of the exposure, and then returns to zenith position for the beginning of the next exposure. Initiating each exposure at zenith produces a series of images in a fixed-declination band that overlap in right ascension. The system begins each exposure at regular right ascension boundaries, ensuring that objects appear in the same location in consecutive night images. Multiple images of the same field over successive nights represent time-series measurements that can be analyzed for transient variability.

TOSS aims to systematically identify optical transient events, and to record these events in a web-based database. Automated post-processing routines (implemented in Matlab®) are used to identify celestial objects in each image, to calculate astrometric and photometric parameters for every detected object, and to store the information in the TOSS database (Table 1). A flowchart of algorithms is shown in Figure 2.

Raw image files are downloaded from each camera to a PC hard drive, and analyzed individually following standard astronomical image processing techniques. Flat-field and dark-frame corrections are applied. A virtual sky is established, and pixels that are significantly brighter than the background are flagged as potentially indicating signal. Object footprints are determined from the bright-pixel map using a two-pass connected-pixel algorithm; centroids are then calculated for each connected object. A point pattern-matching routine is used to correlate observed object centroids to a list of expected objects within the image field, extracted from the Naval Observatory Merged Astrometric Dataset (NOMAD) (Monet et al. 2003; Zacharias et al. 2005).

Astrometry is calculated by least-squares rigid plate alignment to correlated catalog objects. Differential aperture photometry is used to determine raw instrumental magnitudes for each color component, using NOMAD magnitudes for reference star comparison. Derived object parameters are stored in a working database, keyed by the corresponding star catalog ID. Time-series of nightly observations are accumulated for each celestial object in each fixed-declination strip. Each of these steps is described in detail below. To search for transients, individual time-series could then be analyzed for significant deviations from baseline variability.

2. RAW IMAGE PREPARATION

The current TOSS configuration consists of two telescopes on a single mount. Telescopes point at the same meridian, but are offset in declination to produce wider coverage. A 16-inch Meade and a 14-inch Celestron are each attached to Canon 10D cameras. Camera detector arrays are aligned orthogonally with the polar motion. All camera automatic settings are disabled so that no white balance or other gain functions are applied to recorded image data. The common mount orients the telescope array at the overhead meridian for the beginning of every exposure, and then tracks the sky for 90s exposures. At the end of each exposure, the mount returns to the overhead meridian to begin the next exposure. Images are initially stored in the camera memory, and then subsequently transferred to an auxiliary PC hard drive as Canon Raw format (.crw) files which contain raw Bayer array data for each pixel.

Raw .crw files are read directly into Matlab®, using readraw.m, an open-source file that calls dcraw.c (Coffin, 2009). Dcraw.c spatially interpolates the R, G and B values at each Bayer array location to a common grid, and returns a 2048x3072 12-bit red, green and blue frame for each raw image. Each frame is maintained for individual processing. Additionally, a single 'composite grayscale' luminosity frame is also produced from the color components.

Three corrections are applied to each color component of a raw image: a Bad Pixel Mask, a Dark Frame Subtraction, and a Flat Field Correction. A manual list of known bad pixels in the camera pixel array is maintained as an input file. Additional bad pixels are determined from individual dark frame and flat field images. Pixels that are determined to be 'significantly odd' in any color are added to the bad-pixel list. To determine significantly odd pixels in a frame, a median kernel filter is applied to establish a low-spatial-frequency reference frame. For each pixel in the image, the kernel is defined by an asterisk pattern centered at the pixel (Figure 3). The kernel parameters include an inner radius and an outer radius, and only pixels in the asterisk branches between the two radii are included in the kernel for the central pixel. The median asterisk kernel output is unaffected by isolated bad pixels, so only the low spatial frequency components of a frame are passed through the filter. Subtracting the median frame from the original flat field or dark frame removes low-spatial-frequency components, and produces an anomaly frame that ideally should be predominantly flat. Any pixels in the dark frame or flat field images that deviate significantly from the respective anomaly frame are declared as bad pixels. Anomalous pixels are found using Tukey's "inner fence outlier" criteria (Tukey, 1977). The quartiles Q1 and Q3 for the anomaly frame are calculated. If any individual pixel is outside the interval [Q1 − 1.5IQR, Q3 + 1.5IQR] in any color (where IQR = Q3 − Q1), the pixel is declared bad for all colors. Bad pixels detected in dark frame or flat field images are added to the manual list to create a bad-pixel mask for each color component. The bad-pixel mask is applied at appropriate stages in the process.

Once all bad pixels are identified, dark-frame and flat-field corrections are applied, per the procedures described by Berry and Burnell (2005). A master dark frame is created from a series of 90s exposures taken with the shutter closed. A combined flat field is created from a series of twilight frames, exposed to put the frame average at mid-range. A series of flat-field dark frames are acquired with the shutter closed and exposed for the same time as flat fields; these frames are used to create a combined flat-dark frame. The combined flat field is then dark-corrected by subtracting the combined flat-dark frame, producing a master flat-field frame. The master flat field is converted to a master per-pixel gain map by dividing each pixel in the frame by the average value of the central frame region. A bad-pixel map, master dark frame and master gain map are created for each color component.

In the TOSS pipeline, the master dark frame is subtracted from the raw image, with any resulting negative pixels reset to zero. The dark-corrected frame is then divided by the master per-pixel gain map, producing calibrated red, green and blue frames for each raw image. Corrected color frames are used to produce the grayscale luminosity frame (Figure 4) with luminosity defined as the maximum calibrated value of R, G or B at each pixel.

3. OBJECT DETECTION

Celestial objects appear in the calibrated color and grayscale frames. The basic objective of TOSS is to isolate individual objects in an image, and to characterize each object's position and luminosity. Any number of approaches for object detection and characterization are described in the literature (e.g., Andreon et al. 1999), and references cited therein). A crucial step in object detection is background subtraction. To determine a suitable background, the asterisk median filter described above is applied to the grayscale frame. The filter output establishes a low-spatial-frequency reference frame that is used as the estimated background level. The background is subtracted from the grayscale frame to produce an anomaly frame; significantly brighter pixels in the anomaly frame potentially belong to celestial objects. Pixels that are inner fence outliers above the background level are flagged as potential elements of real signal. The TOSS pipeline identifies bright pixels this way and creates a binary frame of pixels that are either significantly above the background (inner-fence outliers, labeled 1) or not (labeled 0). Using the grayscale frame ensures a larger footprint area than if the same method were applied to any single color frame.

In the binary frame, a connected-components algorithm is implemented to identify contiguous groups of bright pixels. The binary image is traversed row-by-row searching for non-zero pixels. The connected-pixel search proceeds across the first row beginning in the upper-left corner of the image. The first non-zero pixel (if any) is labeled as 1. Continuing from pixel to pixel across the row, if another non-zero pixel is encountered, the algorithm checks the pixel to the left; if the left neighbor is already labeled, then the new pixel is given the same label as its left neighbor. If the left neighbor is not labeled, then the most recent label number is incremented by one and the new label value is assigned to the new non-zero pixel.

After the first row, each pixel in subsequent rows is considered after two to four of its eight neighbors have already been assessed and labeled (Figure 5). Suppose the search encounters a non-zero pixel in row $i$, column $j$ of the binary image. Assigning a label to this new non-zero pixel follows a strict order of comparisons. First, if the pixel to the upper left of pixel $(i, j)$ is already labeled, then that label number is assigned to pixel $(i, j)$, and no other adjacent pixels are considered. If the pixel to the upper left is not labeled, then if either the pixel above or the pixel to the left of pixel $(i, j)$ is labeled, but not both, then that label number is assigned to pixel $(i, j)$. If both the pixel above and the pixel to the left of pixel $(i, j)$ are already labeled, and the label is the same, then that label number is assigned to pixel (*i*, *j*). If the two labels are different, then pixel (*i*, *j*) is assigned the same label number as the upper pixel, and a note is made of the equivalence of labels in the two adjacent pixels. If none of the three adjacent pixels already considered has a label, and the pixel to the upper right of pixel (*i*, *j*) is labeled, then that label number is assigned to pixel (*i*, *j*). Finally, if none of the four adjacent pixels has a label, then the most recent label number is incremented by one and that value is assigned to the non-zero pixel (*i*, *j*).

This labeling scheme creates multiple equivalence classes, *i.e.*, a single group of connected pixels could contain multiple labels. A second pass through the labeled objects is required to resolve equivalence classes. The scheme used by TOSS is taken from Press et al. (1992). When the second pass is complete, the binary image is updated to an integer-valued frame: each group of connected pixels is labeled with an integer representing isolated objects; however, background pixels retain their zero designations.

Each group of connected pixels represents a potential celestial object. There is some possibility that a pixel group is not a celestial object but rather an artifact of some kind such as a meteor streak, a satellite track, a cosmic ray trace or an isolated bad pixel that was not previously identified as such. TOSS applies object screening to reject artifacts; however, the use of strict criteria to eliminate all but the most robust objects would be counter to the objectives of TOSS. Similarly, object classification, such as star-galaxy differentiation, is not necessarily required by the TOSS post-processing program. As such, complex object-classification algorithms like those described by Andreon et al. (2000) are beyond the current scope of TOSS. A minimalist approach for suppressing image artifacts is used. Currently, the only gate implemented is object size, which discards groups of connected pixels where the group size is smaller than a pre-determined number of pixels.

The pared list of connected groups is used to establish the locations for each observed object in the frame. The integer-valued frame with object labels is used as a template over the grayscale frame. Corrected luminosity values in the template footprint area of each object are used to establish the centroid position for that object based on the row and column locations in the footprint area:

$$\bar{x} = \frac{1}{n} \cdot \sum_{i=1}^{n} L_{corrected} \cdot column \qquad \bar{y} = \frac{1}{n} \cdot \sum_{i=1}^{n} L_{corrected} \cdot row$$

where $L_{corrected}$ is the calibrated composite grayscale luminosity of the associated pixel. Centroid values are calculated in row and column coordinates. For the image shown in Figure 4, the detected object centroids found by this method are shown in Figure 6.

4. OBJECT IDENTIFICATION

Using centroid locations, each observed object can be identified by comparing the TOSS pattern of objects in a given image with a list of expected sources/objects within the image field of view. Point-pattern matching algorithms are known to give good results for this task. However, several steps are required to pare the expected object list prior to implementing a point-pattern matching routine.  For TOSS, varying atmospheric conditions alter the detection limit from one night to the next, so images from the same field on different nights typically contain a different number of detected objects.  The TOSS pattern-matching routine includes several preparatory steps for accommodating the variance observed in nightly images:

First, an initial 'search' field of view is established, based on zenith pointing at the beginning of each exposure, the image time stamp, and the optical characteristics of the telescope and camera.  Since the telescope is polar aligned, the declination range of the field of view is fixed.  A nominal Right Ascension for the image center is calculated from the exposure start time, converted to local sidereal time (LST):

$$RA = LST - HA$$

Since the frame exposure begins at zenith, the center of the frame when the exposure begins has an Hour Angle (HA) of zero.  Right Ascension in the center of the frame is the same as LST.  TOSS post-processing routines use the algorithm from Curtis (2005) to determine LST from the start-time of the image exposure.  The estimated image field of view is then calculated using a nominal pixel pitch measured from test images.

Pointing jitter necessitates that the search field be slightly larger than the nominal image field of view.  Search field boundaries are used to query the NOMAD catalog for objects below $17^{th}$ magnitude, which is the expected telescope system depth in best observing conditions.  The list of expected objects is typically longer than the list of observed objects in any frame.  The TOSS code pares the expected object list by: (1) refining spatial boundaries of the search field; and (2) decreasing the expected limiting magnitude so that the number of expected objects more closely matches the length of the observed list.

To refine search field boundaries, the location of the observed pattern within the search field is sought.  A spatial pattern match is performed using the brightest few objects in both lists.  The magnitudes for expected objects are taken from the NOMAD catalog.  For observed objects, aperture photometry is used to establish raw apparent magnitudes (details provided in Section 6).  Both lists are then sorted by magnitude.  The spatial pattern of bright observed objects in pixel space is transformed to standard coordinates in a tentative location centered at the middle of the search field.  The observed pattern is then perturbed slightly by small translations and rotations, seeking the best alignment over the set of perturbed positions of the observed pattern.

The observed pattern is converted from pixel coordinates to standard coordinates centered in the middle of the search field.  For a list of observed objects, $\{(x_i, y_i), i=1..n\}$, the coordinate transformation begins by inverting the pixel coordinates:

$x_i' = x_{max} - x_i$      (RA increases to the left, but pixel number increase to the right)

$y_i' = y_{max} - y_i$      (Matlab® numbers the rows top to bottom, from 1 to $y_{max}$)

where $(x_{max}, y_{max})$ are the image dimensions.  Conversion to tentative Standard Coordinates is accomplished by the linear transformations

$$x_i'' = \alpha_0 + \left[x_i' - \frac{1+x_{max}}{2}\right] \cdot \text{Scale}_x \qquad y_i'' = \delta_0 + \left[y_i' - \frac{1+y_{max}}{2}\right] \cdot \text{Scale}_y$$

where (Scale$_x$, Scale$_y$) are the conversion rates between pixels and degrees along the celestial coordinate axes, determined experimentally.

$$\left(\alpha_0 = \frac{\alpha_{min} + \alpha_{max}}{2}, \delta_0 = \frac{\delta_{min} + \delta_{max}}{2}\right) \text{ is the center of the search field.}$$

($\alpha_{min}$, $\alpha_{max}$) and ($\delta_{min}$, $\delta_{max}$) define the search field boundaries.

Rotation through specified perturbation angles $\theta$ is performed about the search field center:

$$\begin{bmatrix} x_i''' \\ y_i''' \end{bmatrix} = \begin{bmatrix} \cos\theta & \sin\theta \\ -\sin\theta & \cos\theta \end{bmatrix} \cdot \begin{bmatrix} x_i'' - \alpha_0 \\ y_i'' - \delta_0 \end{bmatrix} + \begin{bmatrix} \alpha_0 \\ \delta_0 \end{bmatrix}$$

Pattern-match comparisons are made on the Standard Coordinate plane using the same tangent point $(\alpha_0, \delta_0)$ that was used to produce tentative Standard Coordinates for the observed object list. Celestial coordinates for each expected object $(\alpha_i, \delta_i)$ are converted to Standard Coordinates $(X_i, Y_i)$ by:

$$X_i = \frac{\cos(\delta_i) \cdot \sin(\alpha_i - \alpha_0)}{\cos(\delta_0) \cdot \cos(\delta_i) \cdot \cos(\alpha_i - \alpha_0) + \sin(\delta_0) \cdot \sin(\delta_i)}$$

$$Y_i = \frac{\sin(\delta_0) \cdot \cos(\delta_i) \cdot \cos(\alpha_i - \alpha_0) - \cos(\delta_0) \cdot \sin(\delta_i)}{\cos(\delta_0) \cdot \cos(\delta_i) \cdot \cos(\alpha_i - \alpha_0) + \sin(\delta_0) \cdot \sin(\delta_i)}$$

To find the best alignment between observed and expected patterns, the observed pattern is rotated and translated to place a selected observed object over an expected one. Euclidean distances in the Standard Coordinate Frame from each observed object to the nearest-neighbor expected object are calculated. The process is repeated by placing each observed object over every expected object, and re-calculating nearest-neighbor distances. Choices that move any portion of the observed field outside the boundaries of the search field are ignored. The process is repeated for several small rotations, and the sum of nearest-neighbor distances is stored. The global minimum of nearest-neighbor distance sums is taken as the best spatial alignment between the two brightness-paired lists.

The search field is then narrowed to a rectangular frame area that is nominally the same size as the observed field of view, translated and rotated by the same amount as the best spatial alignment. Standard coordinates for the four vertices of the frame area before applying the optimal shift and rotation are:

$$V_1 = \begin{bmatrix} \alpha_0 + \left[1 - \dfrac{1+x_{max}}{2}\right] \cdot \text{Scale}_x \\ \delta_0 + \left[1 - \dfrac{1+y_{max}}{2}\right] \cdot \text{Scale}_y \end{bmatrix} \qquad V_2 = \begin{bmatrix} \alpha_0 + \left[1 - \dfrac{1+x_{max}}{2}\right] \cdot \text{Scale}_x \\ \delta_0 + \left[y_{max} - \dfrac{1+y_{max}}{2}\right] \cdot \text{Scale}_y \end{bmatrix}$$

$$V_3 = \begin{bmatrix} \alpha_0 + \left[x_{max} - \dfrac{1+x_{max}}{2}\right] \cdot \text{Scale}_x \\ \delta_0 + \left[y_{max} - \dfrac{1+y_{max}}{2}\right] \cdot \text{Scale}_y \end{bmatrix} \qquad V_4 = \begin{bmatrix} \alpha_0 + \left[x_{max} - \dfrac{1+x_{max}}{2}\right] \cdot \text{Scale}_x \\ \delta_0 + \left[1 - \dfrac{1+y_{max}}{2}\right] \cdot \text{Scale}_y \end{bmatrix}$$

Then, the optimal rotation and shift are applied to each vertex:

$$V_i' = \begin{bmatrix} \cos(\theta_0) & \sin(\theta_0) \\ -\sin(\theta_0) & \cos(\theta_0) \end{bmatrix} \cdot \left( V_i - \begin{bmatrix} \alpha_0 \\ \delta_0 \end{bmatrix} \right) + \begin{bmatrix} \alpha_0 \\ \delta_0 \end{bmatrix} + \begin{bmatrix} \alpha_{shift} \\ \delta_{shift} \end{bmatrix}, \quad i = 1..4$$

where $\theta_0$ is the optimal rotation angle and ($\alpha_{shift}$, $\delta_{shift}$) is the optimal shift determined by the brute-force spatial alignment routine.

Expected objects in the search field are tested for enclosure within the frame area. For the enclosure test only, standard coordinates for each expected object are rotated about ($\alpha_0$, $\delta_0$) by $\theta_0$:

$$\begin{bmatrix} X_i' \\ Y_i' \end{bmatrix} = \begin{bmatrix} \cos(\theta_0) & \sin(\theta_0) \\ -\sin(\theta_0) & \cos(\theta_0) \end{bmatrix} \cdot \left( \begin{bmatrix} X_i \\ Y_i \end{bmatrix} - \begin{bmatrix} \alpha_0 \\ \delta_0 \end{bmatrix} \right) + \begin{bmatrix} \alpha_0 \\ \delta_0 \end{bmatrix}$$

A simple comparison to the shifted frame area boundaries can then be made to determine enclosure for each expected object.

As the final step for observed object identification, the point-pattern matching algorithm of Murtagh (1992) is employed, which accommodates scaling and translation but requires the two lists to be approximately the same length. The narrowed frame area still typically contains more expected objects than are contained in the observed list because observing conditions are not often optimal, so the expected list is usually deeper. For TOSS, the expected object list, sorted by magnitude, is pared to contain 5% more objects than the observed object list. The list of observed objects is inverted, and rotated about the image center by the optimal spatial alignment:

$$\begin{bmatrix} x_i'' \\ y_i'' \end{bmatrix} = \begin{bmatrix} \cos(\theta_0) & \sin(\theta_0) \\ -\sin(\theta_0) & \cos(\theta_0) \end{bmatrix} \cdot \begin{bmatrix} x_i' - \dfrac{1+x_{max}}{2} \\ y_i' - \dfrac{1+y_{max}}{2} \end{bmatrix} + \begin{bmatrix} \dfrac{1+x_{max}}{2} \\ \dfrac{1+y_{max}}{2} \end{bmatrix}$$

where $x_i' = x_{max} - x_i$ and $y_i' = y_{max} - y_i$ are inverted coordinates. For each object in both lists, the world view of Murtagh is calculated. Nearest-neighbor distances in world-view space provide a robust match between observed and expected objects over the entire observed magnitude range.

Nearest-neighbor distances in world view space also provide a gage for assessing the probability of a good match: smaller world-view distances represent better associations. For TOSS, the spatial nearest-neighbor distance is used as an additional check. If both the world-view and spatial nearest-neighbor distances are small, the association between catalog and observed object is characterized as a good association. Figure 7 gives the catalog association for a number of objects in the raw frame of Figure 4. The TOSS database contains the NOMAD ID for each observed object.

## 5. ASTROMETRY

The current astrometric objectives for TOSS are framed by the desire to provide an accessible database of optical transients that is searchable by location. As an approximation to celestial location TOSS uses linear methods to produce Standard Coordinates for each detected object in an image frame. The use of non-linear astrometric methods, such as described by Jefferys (1987), is less robust to variability noted in TOSS images than standard linear methods, and the added astrometric accuracy is not significant for the goals of TOSS.

Standard Coordinates for TOSS objects correspond to the known locations of the NOMAD catalog objects that have good associations with an observed object. A simple six-constant linear plate transformation is employed, as described by Berry and Burnell (2005), which supports translation, rotation and scaling. The best-fit transformation from pixel coordinates to celestial coordinates is determined from the list of observed object centroids in pixel space and the corresponding celestial coordinates of associated catalog objects. For objects in the observed list that have a valid association with a NOMAD catalog object, the pixel coordinates are first inverted since axis inversion is not supported by the six-constant scheme being employed. The six-constant equations for each object pair are given by:

$$X_i = a \cdot x_i' + b \cdot y_i' + c \qquad Y_i = d \cdot x_i' + e \cdot y_i' + f$$

where $(x_i', y_i')$ are the inverted pixel coordinates of an object centroid, and $(X_i, Y_i)$ are standard coordinates of the corresponding NOMAD catalog object, as determined from their catalog celestial coordinates $(\alpha_i, \delta_i)$ by

$$X_i = \frac{\cos(\delta_i) \cdot \sin(\alpha_i - \alpha_F)}{\cos(\delta_F) \cdot \cos(\delta_i) \cdot \cos(\alpha_i - \alpha_F) + \sin(\delta_F) \cdot \sin(\delta_i)}$$

$$Y_i = \frac{\sin(\delta_F) \cdot \cos(\delta_i) \cdot \cos(\alpha_i - \alpha_F) - \cos(\delta_F) \cdot \sin(\delta_i)}{\cos(\delta_F) \cdot \cos(\delta_i) \cdot \cos(\alpha_i - \alpha_F) + \sin(\delta_F) \cdot \sin(\delta_i)}$$

where

$$\left( \alpha_F = \frac{\alpha_{V_1'} + \alpha_{V_2'} + \alpha_{V_3'} + \alpha_{V_4'}}{4}, \delta_F = \frac{\delta_{V_1'} + \delta_{V_2'} + \delta_{V_3'} + \delta_{V_4'}}{4} \right)$$ is the center of the frame area.

The numerical solution for finding the plate constants a, b, c, d, e and f is over-determined for more than three objects:

$$\begin{bmatrix} x_1' & y_1' & 1 & 0 & 0 & 0 \\ 0 & 0 & 0 & x_1' & y_1' & 1 \\ x_2' & y_2' & 1 & 0 & 0 & 0 \\ 0 & 0 & 0 & x_2' & y_2' & 1 \\ \vdots & \vdots & \vdots & \vdots & \vdots & \vdots \\ x_n' & y_n' & 1 & 0 & 0 & 0 \\ 0 & 0 & 0 & x_n' & y_n' & 1 \end{bmatrix} \cdot \begin{bmatrix} a \\ b \\ c \\ d \\ e \\ f \end{bmatrix} = \begin{bmatrix} X_1 \\ Y_1 \\ X_2 \\ Y_2 \\ \vdots \\ X_n \\ Y_n \end{bmatrix}$$

For TOSS, numerical solution of the over-determined system is taken from Press et al. (1992). Once plate constants are found from the subset of objects with good catalog associations, the transformation is applied to all of the original (inverted) points in the observed objects list, producing Standard Coordinates for each observed object:

$$X_i = a \cdot x_i' + b \cdot y_i' + c \qquad Y_i = d \cdot x_i' + e \cdot y_i' + f$$

Standard Coordinates for each observed object are then converted to Celestial Coordinates by

$$\delta_i = \arcsin\left( \frac{\sin(\delta_F) + Y_i \cdot \cos(\delta_F)}{\sqrt{1 + (X_i)^2 + (Y_i)^2}} \right) \qquad \alpha_i = \alpha_F + \arctan\left( \frac{X_i}{\cos(\delta_F) - Y_i \cdot \sin(\delta_F)} \right)$$

The standard coordinates are listed as the observed object location in the TOSS database.

## 6. PHOTOMETRY

An automated aperture photometry is employed for each color component, following guidelines in Berry and Burnell (2005). Object centroids provide the aperture center for each observed object. The aperture radius is set to 20 pixels as a default value. However, a check is made to ensure that the object footprint lies completely within the aperture as the aperture is increased in cases where the footprint is too large. The annulus used for measuring the background is displaced from the aperture by 4 pixels, creating a gap. When calculating the background, a check is made to determine if the footprint of any adjacent objects lies inside the annulus; if so, those pixels are disregarded when calculating the background level in the outer annulus. Raw apparent magnitudes are calculated for each observed object by

$$m_{apparent} = -2.5 \cdot \log\left( C_{aperture} - n_{aperture} \cdot \left( \frac{C_{annulus}}{n_{annulus}} \right) \right)$$

where $C_{aperture}$ and $C_{annulus}$ are the sums of all the pixel brightness values within the designated aperture areas, and $n_{aperture}$ and $n_{annulus}$ are the total number of pixels in each region. Raw instrumental magnitudes are then determined by identifying three reference stars within the frame. Reference stars are defined to be the brightest three stars in the frame that have NOMAD catalog B, V and R magnitudes. A zero-point uncertainty of the system equivalent to the average difference between the three reference stars and their catalog magnitudes is added to the raw apparent magnitude for each object to obtain raw instrumental magnitude (Boyd 2007):

$$m_{instrument} = m_{apparent} + \left( \frac{1}{3} \sum_{i=1}^{3} \left[ m_{catalog}(ref_i) - m_{apparent}(ref_i) \right] \right)$$

Both raw apparent magnitude and raw instrumental magnitude are produced for each color component as well as for the grayscale luminosity frame. These values are recorded in the TOSS database for each observed object.

Figure 8 shows a histogram of magnitude errors when comparing to the NOMAD catalog. The histogram peaks at zero error across most of the system magnitude range, and, at each magnitude, errors are approximately normally distributed; these characteristics indicate minimal systematic error, and that random sources account for most of the error between instrumental and catalog magnitudes.

7. LIGHT CURVES

Based on the images acquired during a test phase, which were processed with the automated code and analyzed, the TOSS mechanical system and data pipeline have shown to be capable of producing science-grade data. The TOSS system test phase extended from January 10 to March 5, 2010; observations were possible on 29 nights during that period. For each night, between 11 and 31 images were generated within a test field that was centered on the δ-Cepheid AO Aur. The average number of targets detected in each image was approximately 150. All images within the test field were processed with the automated code, and individual .csv files, containing all TOSS database fields listed in Table 1, were created for every frame. Corresponding data from successive nights were then combined to create light curves for individual objects within the test field.

For the generation of light curves, individual .csv files are first assessed and data for suitable objects is written to a single database. Several criteria are used to exclude individual objects from the database:

- If the calculated position of an observed object is more than .00095 degrees from the catalog object to which it is matched, the object is excluded from the database.
- If less than 80% of the objects in a frame are aligned well, then the entire frame is excluded from the database.
- Individual targets were also excluded from the database if any of the component magnitudes are significantly different from the catalog objects to which they are matched.

The cut off values for excluding objects based on magnitude were determined from histograms of differences between observed and catalog magnitudes. The histograms were produced for the root-squared difference between measured and catalog magnitudes for location-matched objects within the observing campaign. A continuous model was fitted to the discrete histogram data, with the following function: $A \cdot x \cdot e^{-B \cdot x}$ The curve peak is at $x = \frac{1}{\sqrt{2B}}$, and the inflection point is at $x = \frac{3}{\sqrt{6B}}$. The distance from the peak to the inflection point is used in the same sense as in a standard normal distribution, *i.e.*, the distance from the peak to the inflection point is exactly one standard deviation. For histogram models of each color, cutoff values are calculated from the continuous curve peak and inflection point as $x = \frac{1}{\sqrt{2B}} + 3.5 \cdot \left(\frac{\sqrt{2}-\sqrt{6}}{2\sqrt{B}}\right)$.

Once all .csv files are processed, light curves for individual objects are extracted from the larger database. The database contains observations of single objects made over multiple nights. Due to observing conditions, not every target will be observed every night. Additional criteria are used to exclude light curves that have obvious errors. An individual observation within a light curve is rejected for several criteria:

- If any individual observation within a light curve is an inner fence outlier (Tukey, 1977), it is omitted from the light curve for all colors.
- If any individual observation in a light curve is more than 0.15 magnitudes away from the average of its two neighbors in any color, it is thrown out of the light curve for all colors.

A known variable, the δ-Cepheid AO Aur is present in the test field and was manually excluded from the magnitude error histogram. A light curve for AO Aur produced by the TOSS pipeline is shown in Figure 9; the shape of the AO Aur light curve is characteristic for δ-Cepheid variables and serves as an example of why we consider our system science-grade data capable. The statistical errors evident in Figure 9 are attributed to the light pollution and increased humidity due to TOSS' current location on the campus of University of California at Santa Barbara.

8. CONCLUSIONS

We have created an automated photometric data pipeline for TOSS that is able to calibrate raw images, identify individual objects within images and cross correlate them with catalog objects as well as perform astrometric and photometric analyses. The automated data-reduction algorithms and results described here have demonstrated the ability of the TOSS system to produce accurate photometric characterizations as a basis for light curves. Additional work in the future will address the photometric characterization of the two commercial cameras used in the system. TOSS is capable of scaling up to a significantly large survey at a modest cost. During the test phase, approximately 500 images were recorded each night by each telescope. Based on an average of 150 detected objects in each image, there are potentially 75,000 light curves that could be extracted from the test data images. While the site chosen for testing was of modest sky quality, it nonetheless allowed us to produce reliable photometry within ~0.1 magnitude. Our pipeline is robust and applicable to a wide variety of photometric datasets and sky surveys.

| 1 | nomad_id | Integer ID taken from the Nomad Catalog. The catalog object corresponds to the nearest neighbor of the detected object in the image |
|---|---|---|
| 2 | nomad_ra | RA in degrees taken from the Nomad Catalog |
| 3 | nomad_dec | Declination in degrees taken from the Nomad Catalog |
| 4 | nomad_gs_mag | Weighted-Average Magnitude, calculated from Nomad Catalog B, V and R magnitudes as gs_mag = 0.2989*red + 0.5870*green + 0.1140*blue |
| 5 | nomad_B_mag | B magnitude taken from the Nomad Catalog |
| 6 | nomad_V_mag | V magnitude taken from the Nomad Catalog |
| 7 | nomad_R_mag | R magnitude taken from the Nomad Catalog |
| 8 | nomad_ref_star | A star in the Nomad Catalog is designated as a Reference Star in the TOSS database if all of the B, V and R magnitudes are present in the Nomad Catalog, and if its gs_mag value is below 15 |
| 9 | centroid_x | Centroid x-value (centroid column) in pixels as calculated by the TOSS code |
| 10 | centroid_y | Centroid y-value (centroid row) in pixels as calculated by the TOSS code |
| 11 | intensity | The sum of all pixel values (ADUs) in the grayscale image over the field of pixels determined to be a part of the detected object |
| 12 | ra | Calculated RA in degrees, determined from plate alignment of the TOSS image with the Nomad Catalog star field |
| 13 | dec | Calculated Declination in degrees, determined from plate alignment of the TOSS image with the Nomad Catalog star field |
| 14 | raw_apparent_mag_gs | Calculated from grayscale intensity as -2.5*log (intensity) |
| 15 | raw_apparent_mag_red | Calculated from red frame intensity |
| 16 | raw_apparent_mag_green | Calculated from green frame intensity |
| 17 | raw_apparent_mag_blue | Calculated from blue frame intensity |
| 18 | raw_instrument_mag_gs | Raw apparent grayscale magnitude scaled to the brightest three reference star grayscale magnitudes in the frame |
| 19 | raw_instrument_mag_red | Raw apparent red magnitude scaled to the brightest three reference star R magnitudes in the frame |
| 20 | raw_instrument_mag_green | Raw apparent green magnitude scaled to the brightest three reference star V magnitudes in the frame |
| 21 | raw_instrument_mag_blue | Raw apparent blue magnitude scaled to the brightest three reference star B magnitudes in the frame |
| 22 | proximity_flag | When the plate constants are applied to detected objects in the image, the final RA and Dec are used to search the Nomad Catalog for the nearest catalog object. When the catalog object is more distant that a threshold value (currently 0.01 degrees), the proximity flag is set to 1; otherwise, it is set to zero. |
| 23 | border_object_flag | When the object centroid is in the border region of the image, this flag is set to 1, otherwise it is set to 0. The border region is currently defined as within 300 pixels of either the left or right edge, or within 200 pixels of the bottom or top edge. |
| 24 | julian_date | Julian date when the image was recorded |
| 25 | file_identifier | A string which represents the date and filename of the image, e.g., 2009_01_28\05_12_321376256.CRW |

**Table 1.** Fields in the TOSS database.

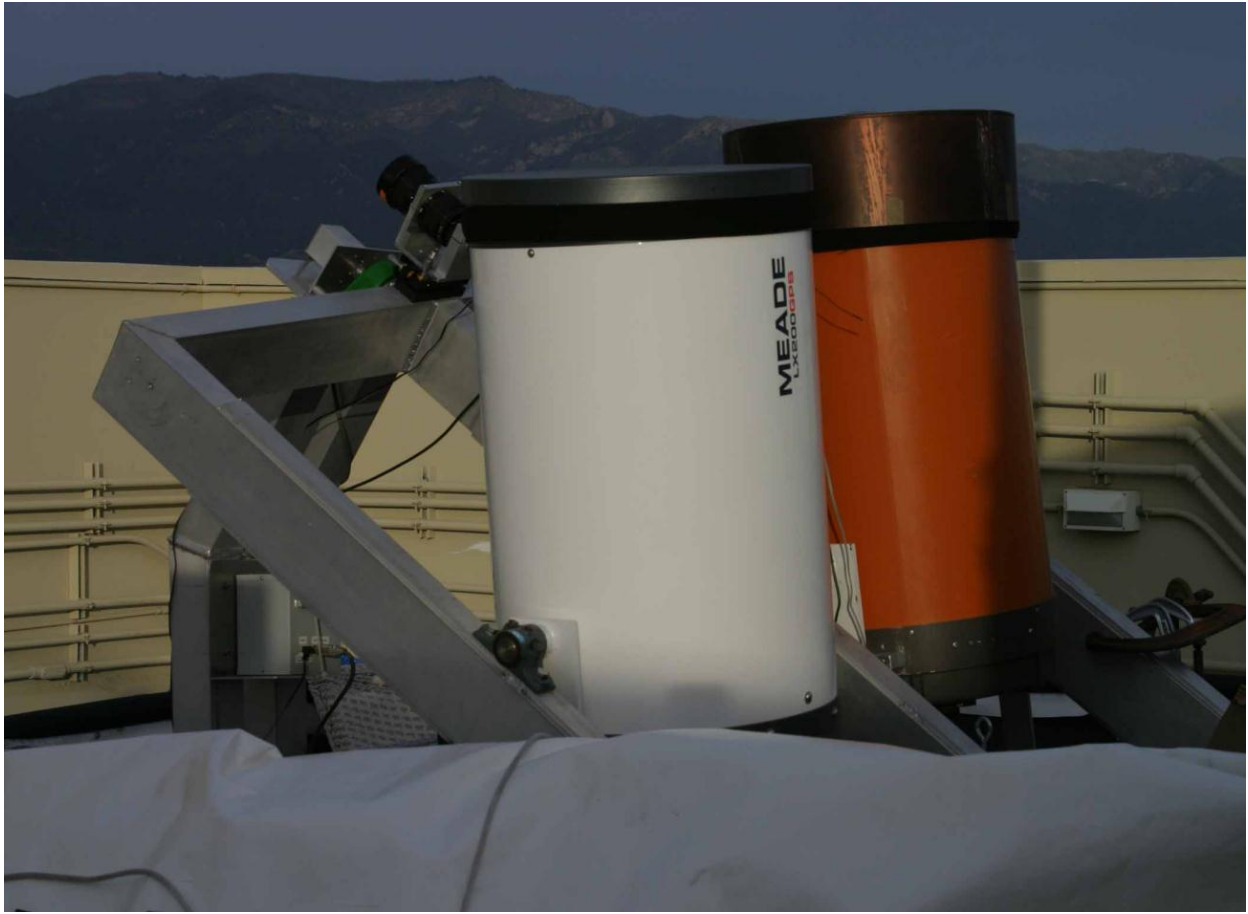

**Figure 1.** The TOSS system of two telescopes on a single semi-equatorial mount, in their home position, pointing at zenith.

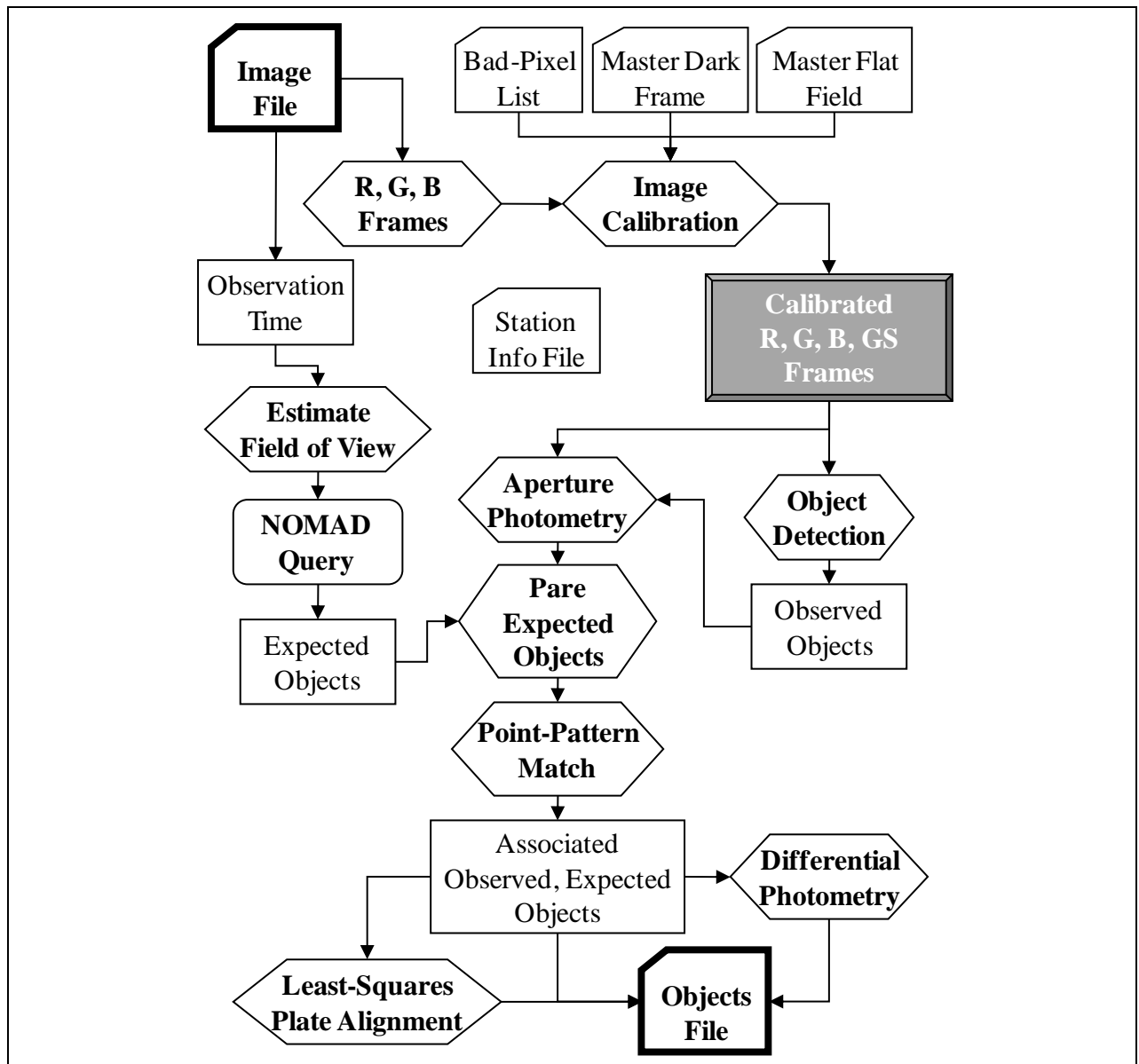

**Figure 2.** The TOSS Post-Processing flow chart, showing the general steps used to extract celestial objects from a single TOSS image. Each image is processed to extract a list of observed objects; the object parameters are written to an Objects (.csv) file that is then added to the TOSS database.

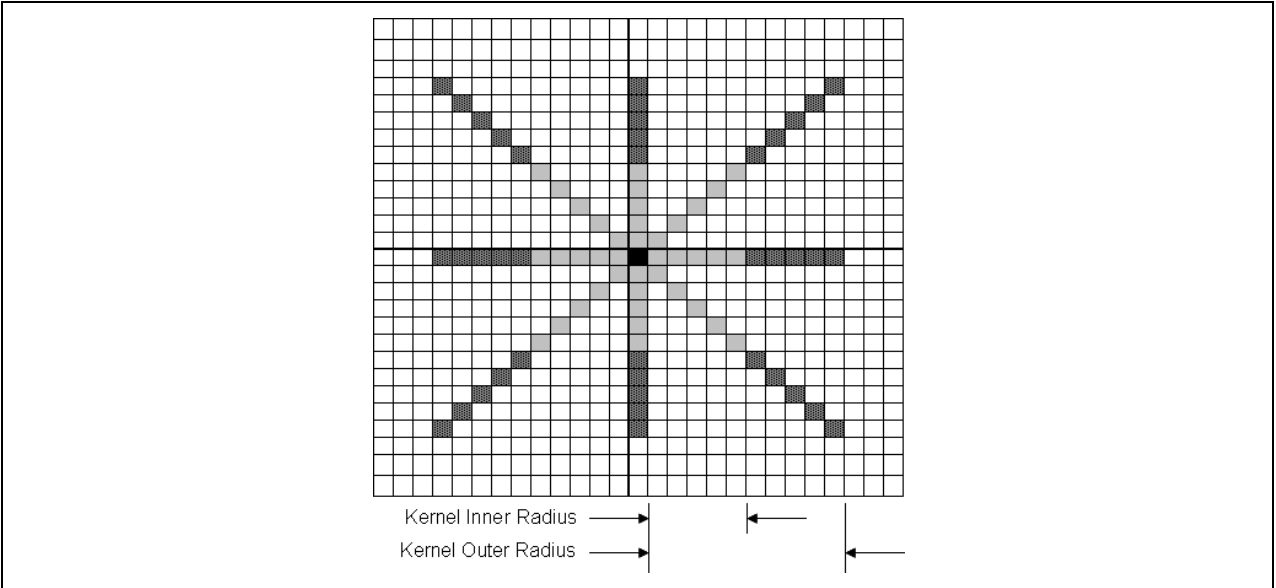

**Figure 3.** The kernel 'asterisk' pattern for establishing a low-spatial-frequency reference frame. For each pixel in the image, the kernel is defined by the inner and outer radii. Pixels in the asterisk branches that are between the inner and outer radii constitute the kernel for the central pixel. The reference value for the central pixel is calculated as the median value of the pixels in the kernel.

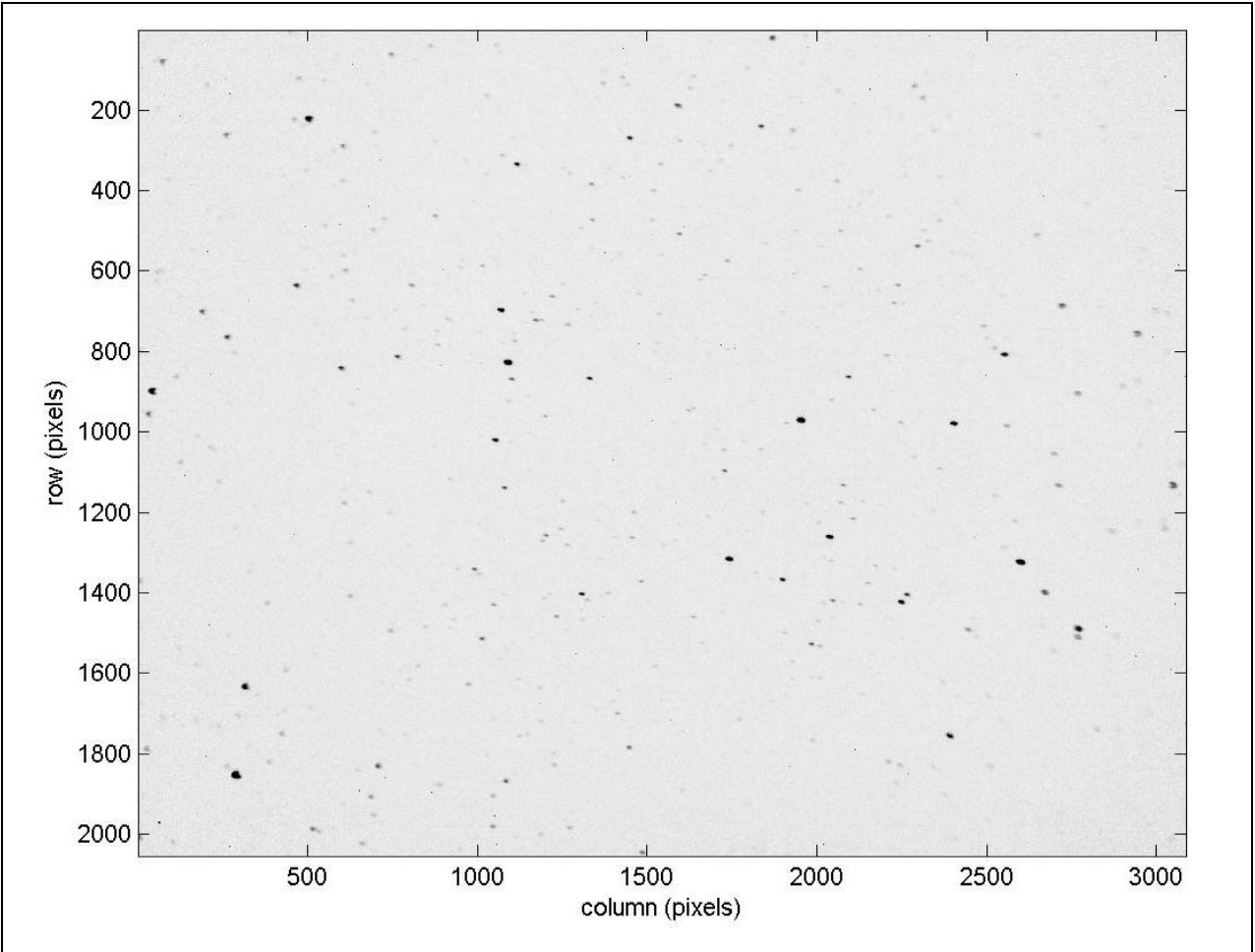

**Figure 4.** A calibrated composite grayscale luminosity image (inverted) after dark-frame and flat-field corrections have been applied.

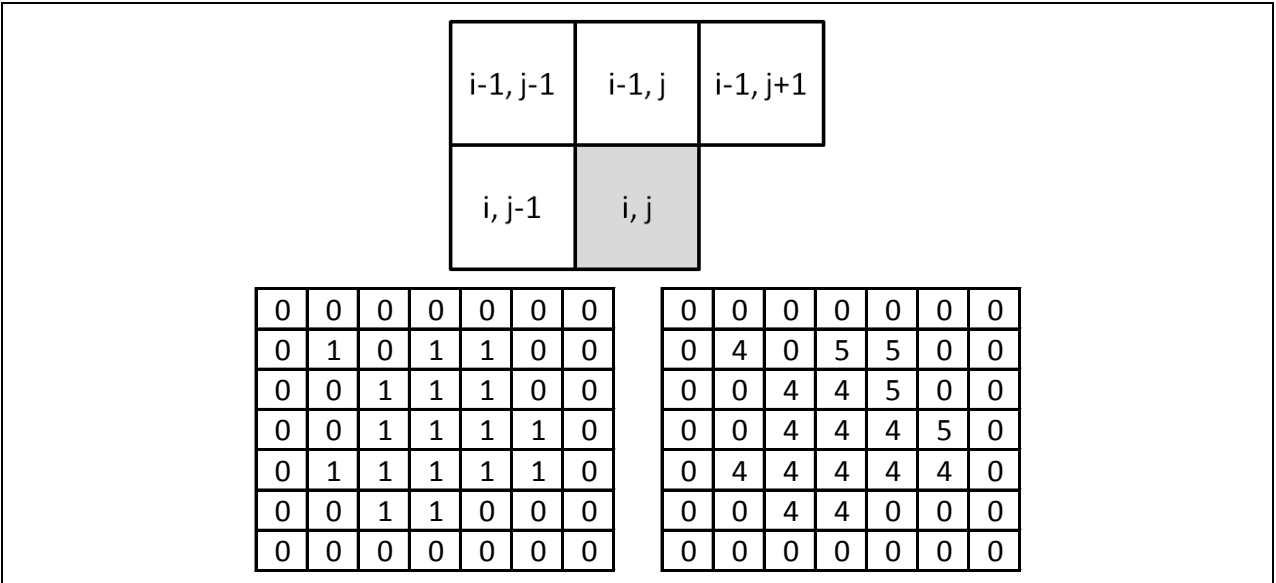

**Figure 5.** Depiction of the connected components algorithm used by TOSS. The image is traversed row-by-row, and left-to-right. When considering the classification of the pixel at row $i$ and column $j$ up to four other pixels have been previously considered. The connected components algorithm labels pixel $(i, j)$ according to the values in the previously-considered pixels. If more than a single label is present in the adjacent pixels, then the equivalence of the labels is noted. If none of the adjacent pixels are labeled, a new label is created and given to pixel $(i, j)$.

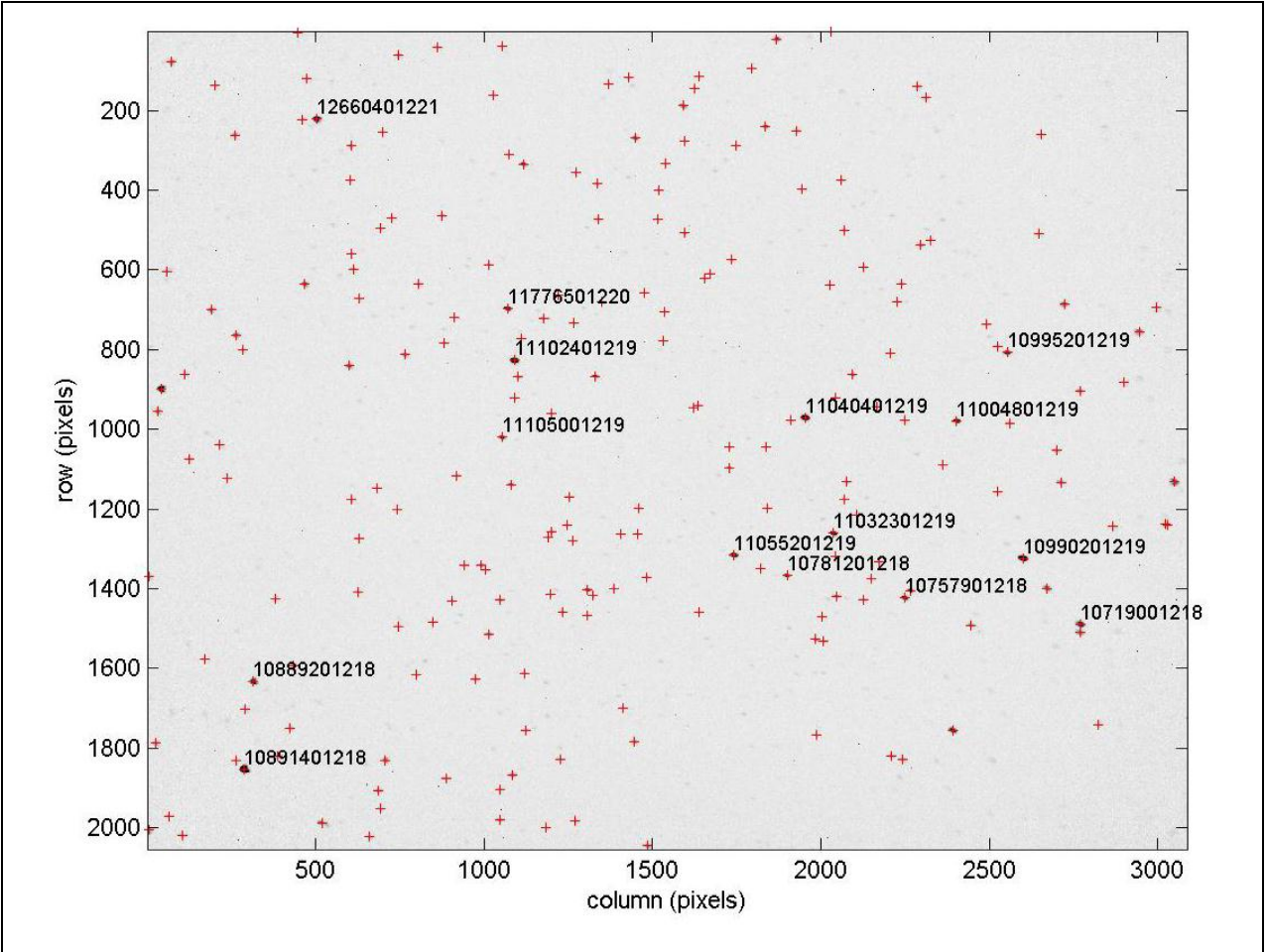

**Figure 6.** Detected-object centroids from the binary image shown in Figure 4, plotted over the original image.

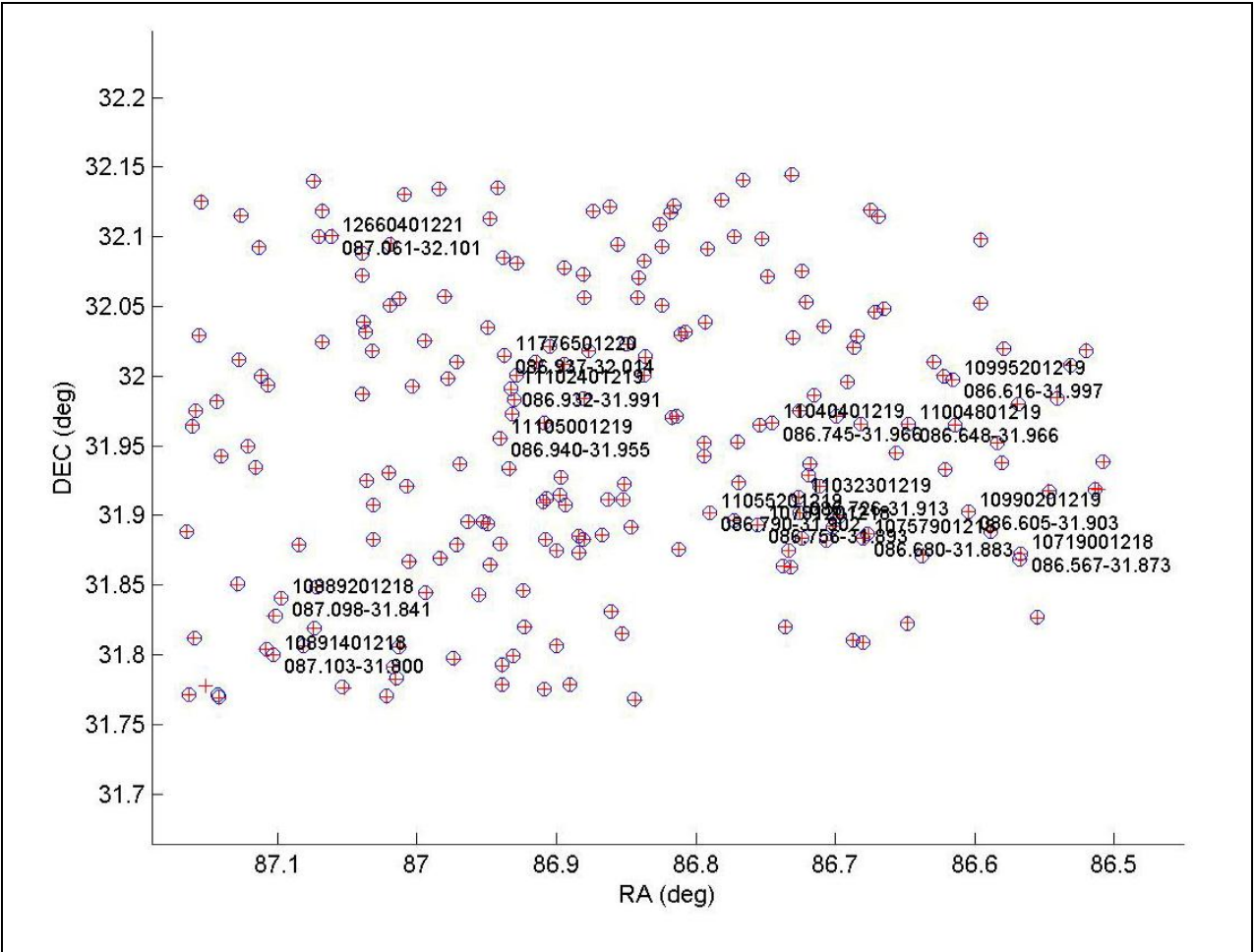

**Figure 7.** Alignment of the observed star field with the NOMAD Star Catalog.

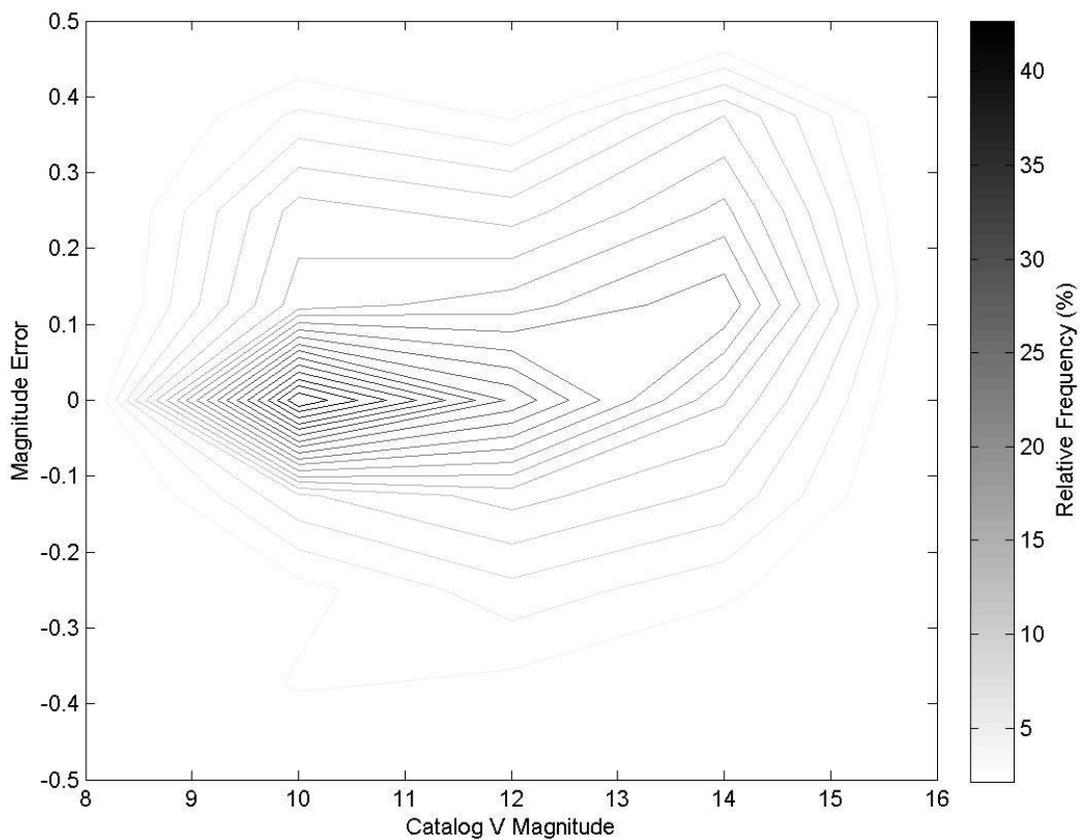

**Figure 8.** Calibration histogram of the TOSS image processing pipeline of the test field, binned by magnitude.

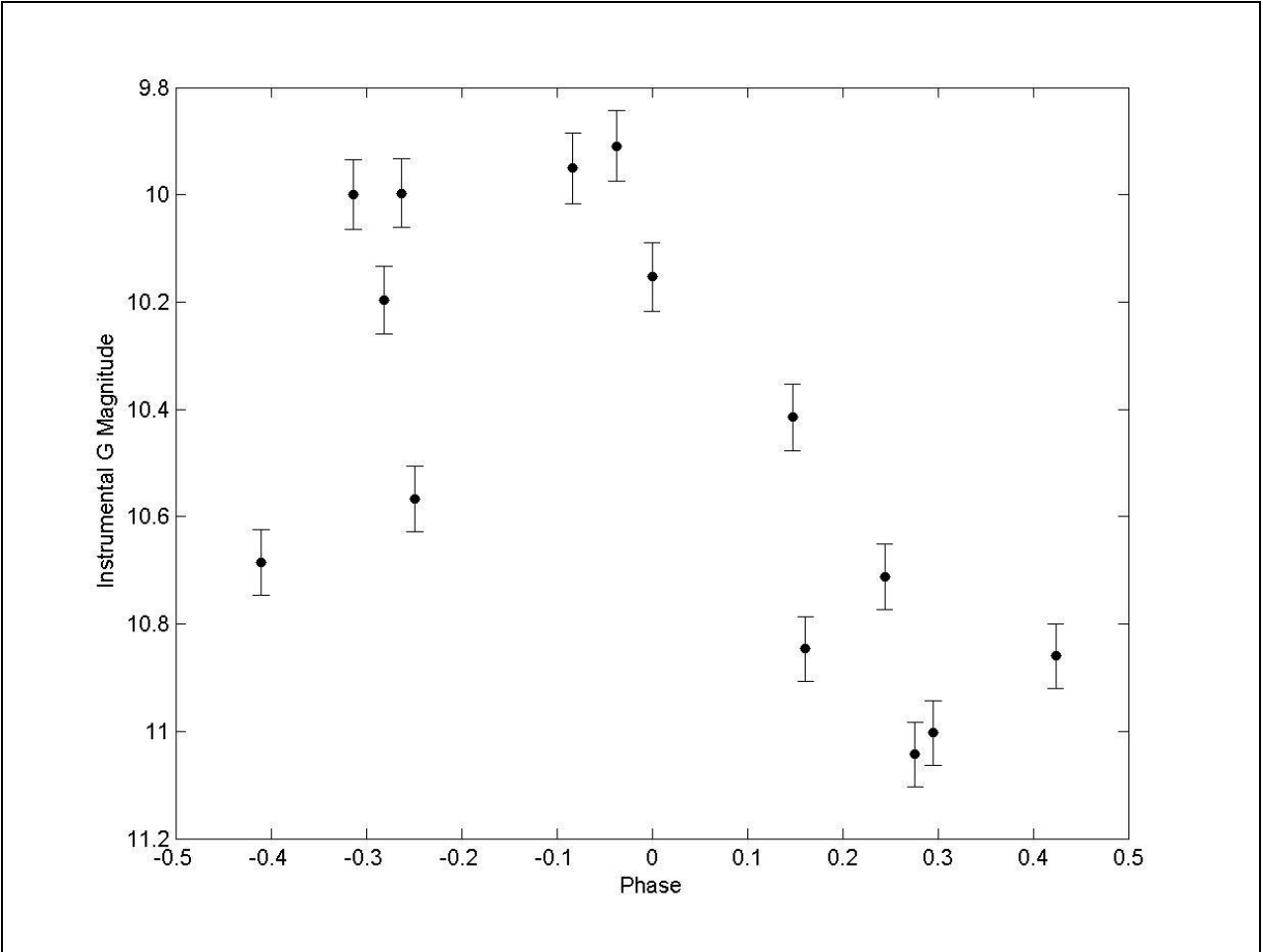

**Figure 9.** Instrumental G magnitude light curve of the δ-Cepheid AO Aur processed using the TOSS automated pipeline. The error bars correspond to the 1-σ rms magnitude error as shown in Figure 8.

REFERENCES

Andreon, S., G. Gargiulo, G. Longo, R. Tagliaferri, and N. Capuano, 2000. Wide field imaging - I. Applications of neural networks to object detection and star/galaxy classification. MNRAS 319, 700-716

Andreon, Stefano, Giorgio Gargiulo, Giuseppe Longo, Roberto Tagliaferri, and Nicola Capuano, 1999. Neural nets and star/galaxy separation in wide field astronomical images, International Joint Conference on Neural Networks, Volume: 6, pp. 3810-3815, Washington, DC.

Berry, Richard, and James Burnell, 2005. The handbook of astronomical image processing, Willmann-Bell, Inc., Richmond, VA, pp. 624.

Boyd, David, 2007. Differential CCD Photometry Using Multiple Comparison Stars, in: Warner, Brian D., Jerry Foote, David A. Kenyon and Dale Mais, editors, Proceedings of the Symposium on Telescope Science, Society for Astronomical Sciences, Big Bear Lake, CA.

Curtis, Howard D., 2005. Orbital Mechanics for Engineering Students, Elsevier Butterworth-Heinemann, Burlington, MA.

Coffin, Dave, Decoding raw digital photos in Linux, source and documentation available online at http://www.cybercom.net/~dcoffin/dcraw/.

Hadjiyska, Elena, Philip Lubin, Scott Taylor and Gary B. Hughes, 2008. Transient Optical Sky Survey Automated Telescope System, Ground-based and Airborne Telescopes II, edited by Larry M. Stepp, Roberto Gilmozzi, Proc. of SPIE Vol. 7012

Jefferys, William H., 1987. Quaternions as Astrometric Plate Constants, AJ 93, 3

Monet, David G., Stephen E. Levine, Blaise Canzian, Harold D. Ables, Alan R. Bird, Conard C. Dahn, Harry H. Guetter, Hugh C. Harris, Arne A. Henden, Sandy K. Leggett, Harold F. Levison, Christian B. Luginbuhl, Joan Martini, Alice K. B. Monet, Jeffrey A. Munn, Jeffrey

R. Pier, Albert R. Rhodes, Betty Riepe, Stephen Sell, Ronald C. Stone, Frederick J. Vrba, Richard L. Walker, and Gart Westerhout, 2003. The USNO-B Catalog, AJ 125:984-993

Murtagh, F., 1992. A New Approach to Point-Pattern Matching. PASP 104:301-307

Press, W.H., S.A. Teukolsky, W.T. Vetterling, and B.P. Flannery, 1992. Numerical Recipes in C: The Art of Scientific Computing, 2d Edition, Cambridge University Press, New York.

Paczyński, Bohdan, 2000. Monitoring All Sky for Variability, PASP 112, 1281-1283, arXiv:astro-ph/0005284v2.

Pojmański, G., 1997. The All Sky Automated Survey, Acta Astronomica, 47, 467, arXiv:astro-ph:9712146.

Tukey, John W., 1977. Exploratory Data Analysis, Addison-Wesley, Reading, MA.

Zacharias,N., Monet,D.G., Levine,S.E., Urban,S.E., Gaume,R, & Wycoff,G.L., 2005. The Naval Observatory Merged Astrometric Dataset (NOMAD), AAS 205, San Diego meeting, poster. Database is available online from the US Naval Observatory Image and Catalog Archive at http://www.nofs.navy.mil/data/fchpix/.